\documentclass[11pt]{article}

\usepackage{upgreek}
\usepackage[utf8]{inputenc}
\usepackage[english]{babel}
\usepackage{amsmath,amsfonts,amssymb}
\usepackage{mathpazo}  
\usepackage[scaled]{helvet}
\usepackage[T1]{fontenc}
\usepackage{url}
\usepackage[colorlinks=true, allcolors=blue]{hyperref}
\usepackage{graphicx,xcolor}
\usepackage[left=48pt, right=42pt, top=46pt, bottom=60pt, headheight=15pt, headsep=10pt, letterpaper]{geometry}
\usepackage{siunitx}

\begin{document}

\title{Shot noise limited high speed stimulated Raman microscopy}
\author{Xavier Audier\textsuperscript{1}, Nicolas Forget\textsuperscript{2}, Hervé Rigneault\textsuperscript{1,*}}

\maketitle

\begin{center}
\textsuperscript{1} Aix Marseille Univ, CNRS, Centrale Marseille, Institut Fresnel, 13013 Marseille, France \\
\textsuperscript{2} FASTLITE, 06560 Valbonne, France \\
\textsuperscript{*} Corresponding author: herve.rigneault@fresnel.fr 
\end{center}

\begin{abstract}
We report a shot noise limited high speed stimulated Raman microscopy platform allowing to acquire molecular vibrational spectra over \SI{200}{\per \cm} in \SI{12}{\micro \second} at a scan rate of 40kHz. Using spectral focusing together with optimized acousto-optics programmable dispersive filters the designed low noise imaging platform performs chemical imaging of dynamical processes such as Mannitol crystal hydratation and reaches a signal to noise ratio sufficient to perform label free histological imaging on frozen human colon tissue slides.
\end{abstract}

\section{Introduction}

Stimulated Raman scattering (SRS)~\cite{Bloembergen_1967} imaging has gain tremendous interest over the last decade due to its ability to perform label free chemical imaging in biological sample~\cite{Cheng2015,Zhang2018}. In SRS microscopy two laser pulses, a pump at frequency $\omega_p$ and a Stokes at frequency $\omega_s$, are focused on a sample to generate an image by point scanning. If the frequency difference $\omega_p - \omega_s$ equals a molecular vibrational frequency $\Omega_R$ an energy transfer occurs between the pump and the Stokes beam~\cite{Rigneault2018} that can be detected using dedicated modulation schemes.  Since SRS imaging first demonstrations~\cite{Freudiger08_science,Nandakumar09_njp}, efforts have been pursue to perform fast images~\cite{Saar03122010}, but also to acquire a complete vibrational spectrum at each pixel~\cite{Seto2013,Rock2013,Liao2015,Liao2016,He2017,Kumar2018,Figueroa2018,Ragni2019,Ozeki2019}. The fastest approaches to acquire a full spectrum~\cite{Liao2016,He2017} use the SRS spectral focusing scheme that allow to retrieve spectral resolution by adding dispersion on the femtosecond pulses~\cite{Gershgoren2003, Hellerer2004}. When the pump and Stokes pulses are identically chirped, a narrow wavenumber range can be probed within the spectral bandwidth allowed by the pulses (fig.~\ref{fig:figure1}d). Not only is spectral resolution recovered, but the probed wavenumber can easily be tuned by adjusting the time delay between the two pulses, allowing for spectral scanning without the need to change the lasers center wavelengths or fast spectrometer. Recently, acousto-optic programmable dispersive filters (AOPDF) have been used as ultrafast delay lines in different applications, such as transient absorption and terahertz spectroscopy~\cite{Urbanek2016,Audier2017}. Contemporary to this work, spectral focusing and AOPDF has been combined to rapidly sweep the delay between two chirped femtosecond pulses in an SRS microscope~\cite{Alshaykh2017}. Raman spectra covering the full lipid band (\SIrange[range-units = single]{2800}{3050}{\per \centi \meter}) were acquired at \SI{33}{\micro \second} per pixel, with a \SI{25}{\per \centi \meter} resolution. Although appealing, this demonstration were done on artificial samples providing strong Raman signals. In particular, the sensitivity was not showcased on biological tissues, and the increased acquisition rate was not applied to the study of dynamic behaviors, while those two aspects are essential for applications.

Here, we report hyperspectral SRS imaging over \SI{200}{\per \cm} in \SI{12}{\micro \second} at a scan rate of 40kHz with shot noise limited sensitivity. By characterizing the noise in our system and engineering the optimal laser pulses, we have designed an SRS imaging platform combining spectral focusing together with a faster AOPDF delay line to surpass the acquisition speed and sensitivity demonstrated previously. We demonstrate optimized speed and sensitivity by targeting two major applications that are dynamic chemical imaging and label-free histology on human tissue samples.

\section{Setup Description}
The SRS pump and Stokes beams are generated by a commercial femtosecond laser system (Chameleon OPO-VIS, Coherent) working at \SI{80}{\mega \hertz} repetition rate. (fig. \ref{fig:figure1}.a)
The pump (\SI{800}{\nano \meter}) and Stokes (\SI{1045}{\nano \meter}) optical pulses are synchronized, and both can be modeled as Gaussian pulses with temporal full widths at half maximum of \SI{160}{\femto \second}.
The pump laser is sent to an AOPDF (HR or WB Dazzler, Fastlite) which acts as both a tunable dispersive medium and an ultra fast delay line (see details below).
The Stokes laser is sent through an acousto-optic modulator (MT200-A0.2-1064, AA optoelectronics) driven by a sinusoidal modulation at one fourth of the laser repetition rate (\SI{20}{\mega \hertz}).
The first diffraction order of the Stokes beam is collected and sent through a double pass grating pair (custom gratings, Wasatch Photonics) where it undergoes negative dispersion.
The modulated and negatively chirped Stokes beam is then recombined with the pump beam by means of a dichroic mirror before being sent to an inverted scanning microscope (TiU, Nikon).
The two lasers are focused in the sample using a 20x air objective (0.75NA, CFI Plan Apo Lambda, Nikon) and collected with the same objective in the forward direction.
The Stokes beam is filtered out using an optical short pass filter (FES0900, Thorlabs), and the pump is collected using a photodiode who's output is then fed to a lock-in amplifier module.
The photodiode, the lock-in amplifier and frequency divider are commercial systems optimized to work at \SI{20}{\mega \hertz} (SRS Lockin Module, APE).
The lock-in amplifier bandwidth was reduced to \SI{1.35}{\mega \hertz} using an electronic lowpass filter (EF508 Thorlabs, 5\textsuperscript{th} order).
The signal from the lock-in was sent to a data acquisition card (ATS460, AlazarTech) and acquired with a sampling rate of \SI{20}{\mega \hertz}.
Simultaneously to SRS, and to address biological applications, second harmonic generation (SHG) was recorded at \SI{400}{\nano \meter} in the epi direction, using  a dichroic mirror (770dcxr, Chroma) band-pass filter (HQ400/40, Chroma), and photomultiplier tube (R9110 tube and C7950 data socket, Hamamatsu).
Average laser powers at the sample plane were \SIlist[list-final-separator = {and}]{15;20}{\milli \watt} for the pump and Stokes beams, respectively.

\begin{table}
\centering
\begin{tabular}{r|c|c|c}
Model & Rep. rate (\si{\kilo \hertz}) & Range (\si{\pico \second}) & Ratio (\si{\femto \second / \micro \second})\\
\hline
HR & 30.6 & 8.5 & 260\\
\hline
WB & 40 & 3.5 & 161\\
\hline
\end{tabular}
\caption[AOPDF characteristics]{Characteristics of the two AOPDF types. Both models consist of a 25-mm-long $TeO_2$ crystal, but the cut angle is different for each, to allow for different acoustic propagation speed. The repetition rate is the maximum rate at which successive acoustic wave can be sent inside the crystal. The range is the maximum delay allowed by the optical index difference between the fast and slow axis, the values are given for \SI{800}{\nano \meter} light. The time ratio gives the amount of delay added per microsecond of acoustic propagation.}
\label{tab:AOPDFspecs}
\end{table}

\section{AOPDF Description}
\begin{figure}[htbp]
\centering
\includegraphics[scale=1]{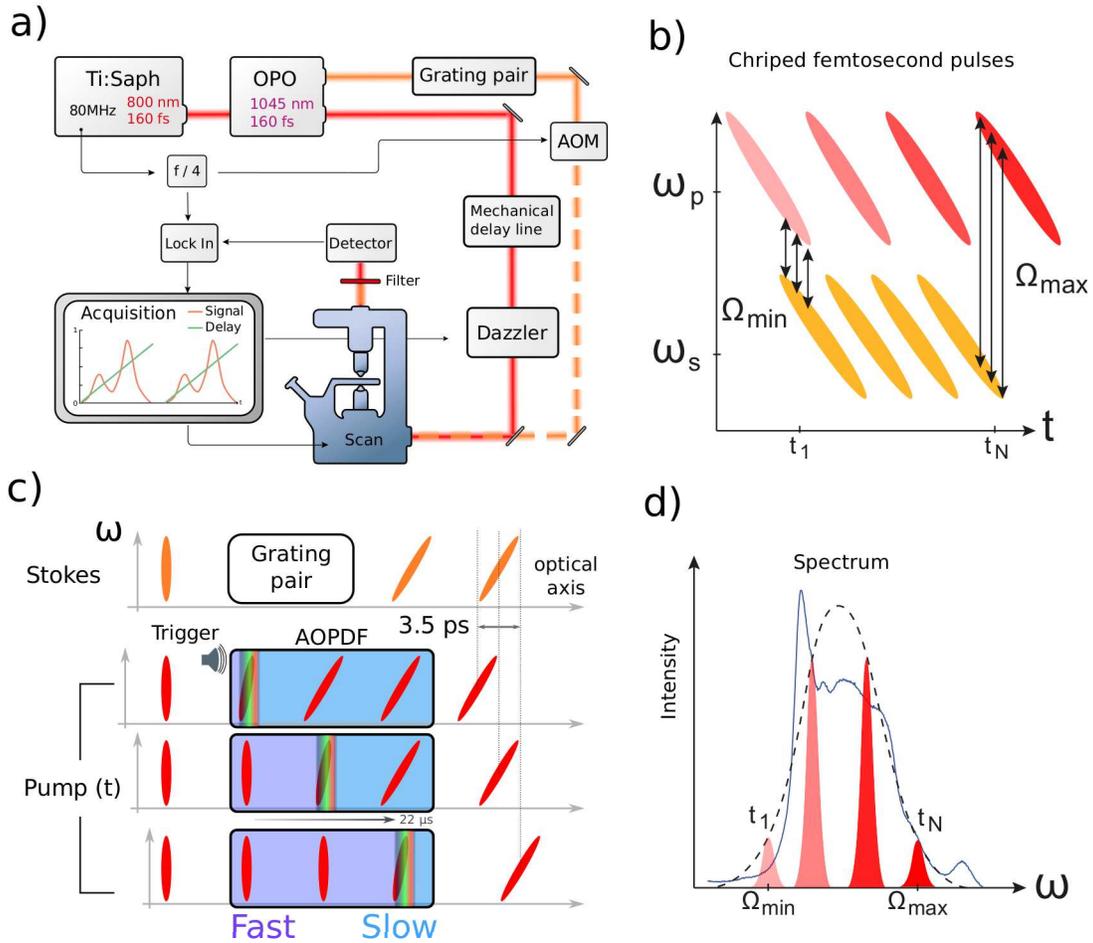}
\caption{a) Setup schematics, OPO: optical parametric oscillator, AOM: acousto-optic modulator. b) Spectral focusing scheme: the instantaneous frequency difference between the pulses is a function of the pulses relative delay. c) Acousto-optic programmable dispersive filter (AOPDF): the acoustic wave propagating inside the crystal imprints both negative dispersion, and a delay that changes linearly with the acoustic wave propagation time. d) Signal and resolution obtained by using chirped pulses that are delayed as in b). }
\label{fig:figure1}
\end{figure}

The AOPDF consists of a birefriengent crystal in which an acoustic shearing wave co-propagates with an ordinary-polarized optical beam (fast axis).
Each acoustic frequency interacts with a specific optical frequency (acousto-optic phase-matching relationship) and gives rise to an extraordinary-polarized diffracted beam (slow axis)
Therefore, using the proper acoustic waveform, one can imprint an arbitrary phase profile on incoming optical pulses~\cite{Verluise2000}.
In addition to this tunable phase profile, the propagation of the acoustic wave inside the crystal effectively imprints a delay on successive optical pulses that increases linearly over time (fig.~\ref{fig:figure1}.c).
Using both of these properties, one can address the two requirements for spectral focusing SRS: chirped optical pulses and fast delay scanning.
Two different AOPDF, the High-Resolution (HR) and Wide-Band (WB), have been used in this study.
They are based on the same concept, but have different working parameters, reported in table~\ref{tab:AOPDFspecs}.
In particular, the WB model has a higher repetition rate, a higher diffraction efficiency, and a shorter delay range, making it better suited for vibrational spectroscopy.

\section{Noise levels}
\begin{figure}[htbp]
\centering
\fbox{\includegraphics[width=\linewidth]{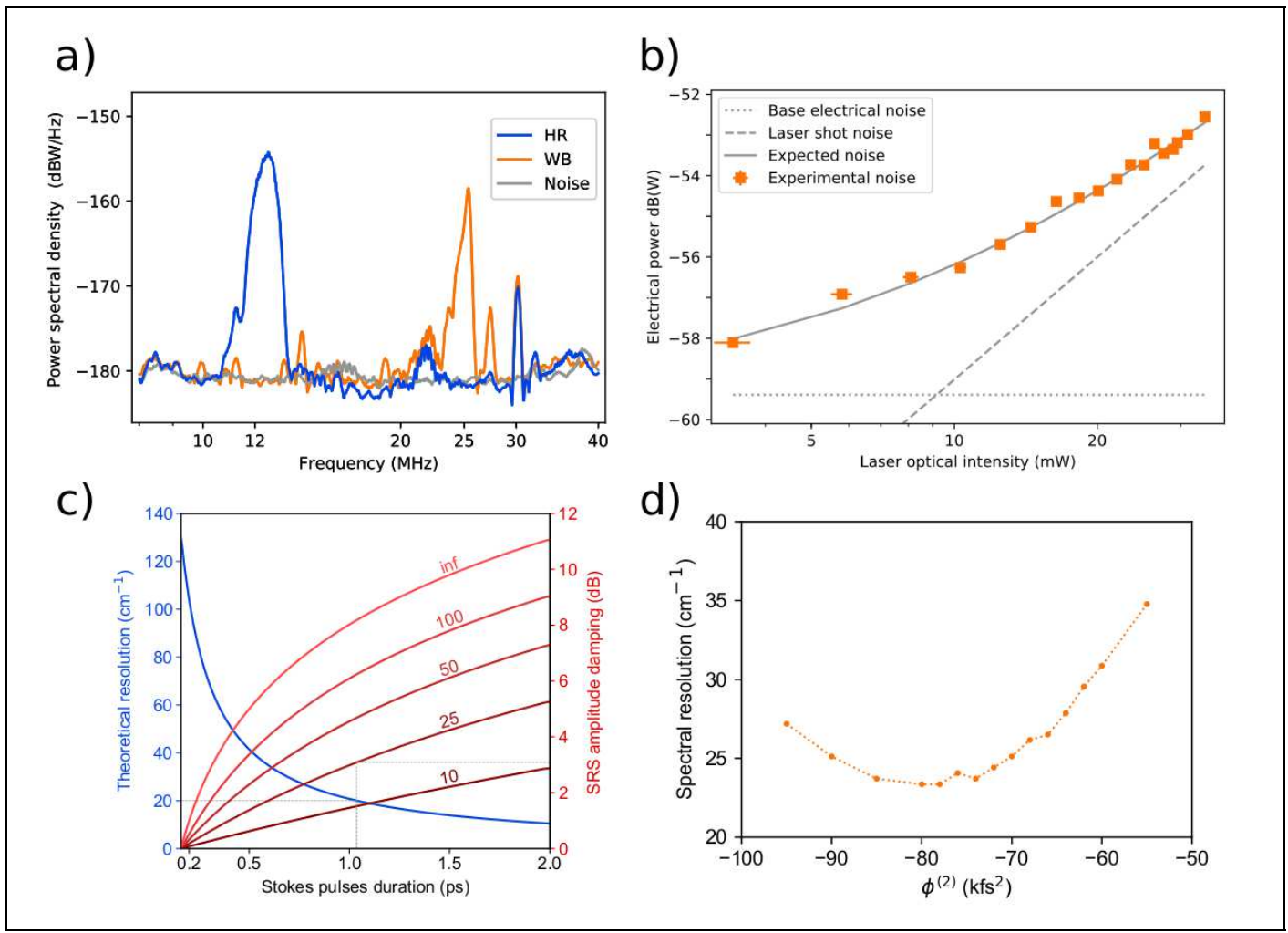}}
\caption{a) Laser intensity noise for the HR and WB AOPDF. b) System output noise power compared to shot noise and electronic noise, for increasing laser power. c) SRS signal loss and resolution gain as a function of the chirped Stokes pulse duration. The different losses (red lines) corresponds to different linewidths for the probed molecular vibration, indicated in red in \si{\per \cm}. d) Achieved spectral resolution as a function of the dispersion (second order phase) applied using the AOPDF.}
\label{fig:figure2}
\end{figure}

The laser system used in this study is shot noise limited for electronic frequencies around \SI{20}{\mega \hertz} and average laser intensities at least up to \SI{70}{\milli \watt}~\cite{AudierNoise2019}.
However, to investigate if the AOPDF introduces additional noise, we measure the laser noise in the presence of the AOPDF, at the detector plane, using \SI{5}{\milli \watt} of average laser power and a commercial photodiode (Det10A, \SI{350}{\mega \hertz} bandwidth, Thorlabs).
The photodiode was loaded with a \SI{50}{\ohm} resistor and the resulting voltage was filtered with a \SI{5.6}{\mega \hertz} high-pass filter (EF515, 6\textsuperscript{th} order, Thorlabs).
The purpose of this filter is to damp the repetition rate and harmonics of the AOPDF~(Table~\ref{tab:AOPDFspecs}).
The filtered photodiode output was sent to a spectrum analyzer (HF2LI, Zurich instrument) and its electrical power spectral density was acquired between \SIlist[list-final-separator = {and}]{8;40}{\mega \hertz} (fig~\ref{fig:figure2}a)).

The power spectral density associated with shot noise is estimated at \SI{-191}{\dB \watt \per \hertz}, which is below the detection limit of our spectrum analyzer (\SI{-180}{\dB \watt \per \hertz}).
However, the laser intensity noise introduced by the AOPDF was high enough to be measured without amplification.
The laser intensity noise is particularly high around \SI{12}{\mega \hertz} for the HR AOPDF, and around \SI{25}{\mega \hertz} for the WB AOPDF.
This point has not been addressed in previous studies, and may have limited the sensitivity of previous systems using these devices.
It is suspected that the amplitude noise imprinted on the laser by the AOPDF results from the interference of successive acoustic waves by reflection and diffusion inside the crystal.
Consistent with this hypothesis, adding a waiting time between successive acoustic pulses significantly decreased the noise level (data not shown).

It is possible to avoid the excess noise introduced by the AOPDF by selecting a suitable modulation frequency and lock-in bandwidth.
Using the lock-in system described previously, with a modulation frequency of \SI{20}{\mega \hertz} and a \SI{1.35}{\mega \hertz} bandwidth, the measured noise from this system matches exactly the expected shot noise from the laser (fig~\ref{fig:figure2}b)).
For average laser intensities above \SI{9}{\milli \watt}, detector electronic noise becomes negligible compared to the laser shot noise.

\section{Resolution and signal level}
Spectral focusing is used to recover spectral resolution when performing SRS with two femtosecond pulses, as illustrated in figure~\ref{fig:figure1}b.
The spectral resolution that can be achieved using spectral focusing increases with the amount of dispersion added to the two pulses, while the SRS signal drops (fig~\ref{fig:figure2}.c)
The associated mathematical derivations can be found in \cite{Su2013}.
We will focus here on the lipid vibrational band between \SI{2850}{\per \centi \meter} and \SI{3000}{\per \centi \meter} where spectral features are typically larger than \SI{20}{\per \centi \meter}.
For this reason, both pump and Stokes pulses were chirped to \SI{1}{\pico \second}, by adding negative dispersion with the AOPDF and grating pair, respectively.
Such pulse dispersion corresponds to a theoretical value of \SI{22}{\per \centi \meter} for the SRS spectral resolution, which is confirmed experimentally using the Raman line of DMSO at \SI{2913}{\per \centi \meter} (fig~\ref{fig:figure2}.d).
Because the AOPDF allows for tunable dispersion, the amount of added second order phase was optimized (fig~\ref{fig:figure2}.d), and the resulting total negative dispersion is consistent with the optics in the system.
The \SI{-77}{k\femto \second \square} second order phase added by the AOPDF includes the \SI{-59}{k\femto \second \square} required to negatively chirp the pulses at the sample plane, the \SI{-13}{k\femto \second \square} required to compensate the dispersion introduced by the AOPDF crystal, and the \SI{-5}{k\femto \second \square} compensating the rest of the optics on the optical path.
The expected drop in SRS signal resulting from the use of chirped pulses depends on the linewidth of the probed molecular bond (fig~\ref{fig:figure2}.c).
In the lipid band we can expect linewidths of \SI{25}{\per \centi \meter}, and as a result a \SI{3}{\dB} drop in signal.

\section{Chemical imaging of multiple species}

\begin{figure}[htbp]
\centering
\fbox{\includegraphics[scale=0.7]{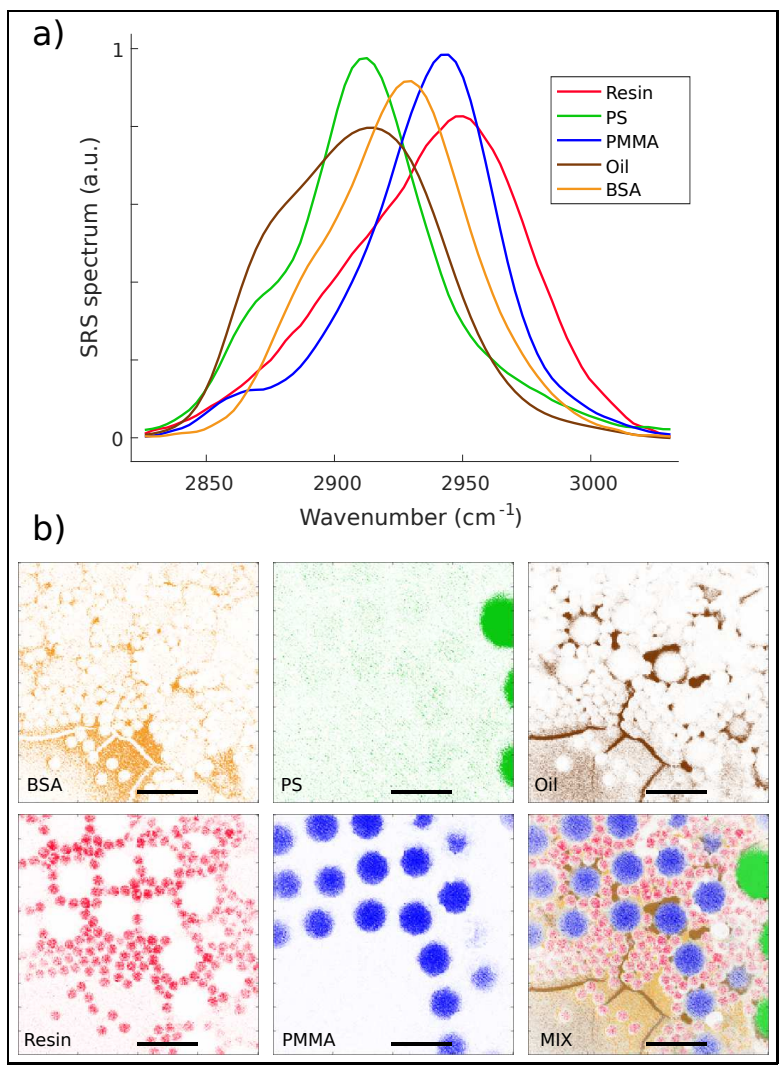}}
\caption{a) SRS spectra of pure species acquired with the AOPDF delay line (Resin: melanine, PS: polystyrene, PMMA: polymethylmethacrylate, Oil: olive oil, BSA: bovine serum albumin). b) Image components associated with each pure species, as extracted from the hyperspectral image. MIX represents the summed image of all components. Pixel dwell time \SI{25}{\micro \second}, scale bar: 25um, total image acquisition time: 1s}
\label{fig:figure4}
\end{figure}

Figure~\ref{fig:figure4}.a shows the SRS spectra acquired for 5 pure species:  melamine (Resin, bead 5um), polystyrene (PS, bead 20um), polymethylmethacrylate (PMMA, bead 10um), bovine serum albumin (BSA) crystals an olive oil solution. An artificial sample composed of these 5 species was prepared and imaged using the WB AOPDF using a pixel dwell time of \SI{25}{\micro \second} that allowed to record the full Raman spectrum between \SI{2850}{\per \centi \meter} and \SI{3000}{\per \centi \meter} with a resolution $\approx$\SI{22}{\per \centi \meter}. The recorded hyperspectral image was projected over the known pure spectra using linear decomposition that allowed to map the 5 species fig~\ref{fig:figure4}.b. The different components are easily distinguishable without further averaging.

\section{Dynamic imaging of chemical reactions}
We use here our fast SRS imaging platform to monitor the dynamic of chemical reactions. We concentrate on Mannitol, a common excipient used in pharmaceutical industry, that can crystallize in several polymorphic forms: $\alpha$, $\beta$, and $\delta$.
The $\delta$ polymorph can change into $\beta$ upon hydration, and both forms can be identified through their Raman spectra in the lipid band.

\begin{figure}[htbp]
\centering
\fbox{\includegraphics[scale=1]{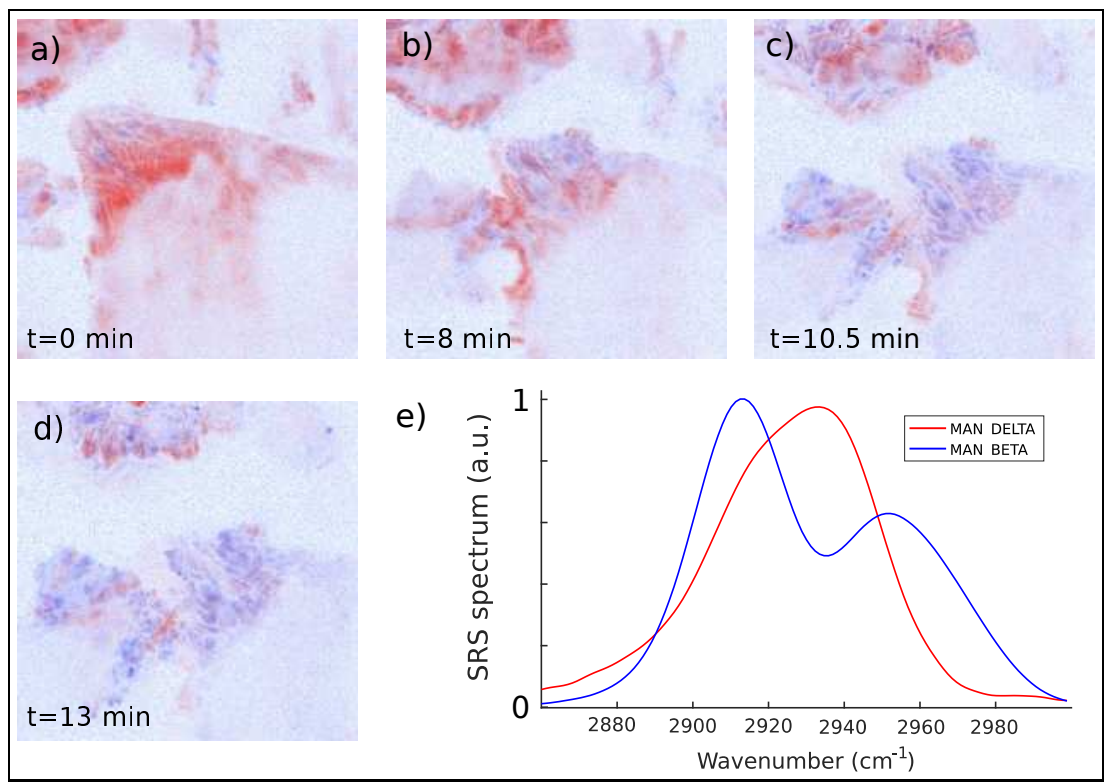}}
\caption{a)-d) Timelapse of a $\delta$-Mannitol (red) crystal transforming into $\beta$-Mannitol (blue) in the contact of water vapour. The field of view is 100 micron large, pixel dwell time \SI{40}{\micro \second}, e) SRS spectra of pure $\delta$-Mannitol (red) and $\beta$-Mannitol (blue).}
\label{fig:Mannitol}
\end{figure}

Pure $\delta$ Mannitol crystals were prepared between a microscope slide and a coverslip and imaged over twenty minutes during the introduction of water vapor. Snapshots of the user interface during the acquisition are displayed in figure~\ref{fig:Mannitol} (a-d), while the full video recording is available (Visualization 1). $\delta$ Mannitol (red) transforms in $\beta$ Mannitol (blue) over time upon water hydratation. We used the HR AOPDF with a frame rate of 1 image every \SI{1.6}{second} (two successive images were averaged to increase the signal to noise ratio).
The hyperspectral images were processed on the fly by projecting them on the components corresponding to pure $\beta$ and $\delta$ Mannitol (fig~\ref{fig:Mannitol}.e), therefore providing live feedback on the crystal transformation process.
Because of the dynamic nature of the process, image refocusing and tracking of the relevant feature was necessary, and was made possible with live two color rendering.

\section{Label-free histological recordings}
\begin{figure}[htbp]
\centering
\fbox{\includegraphics[width=\linewidth]{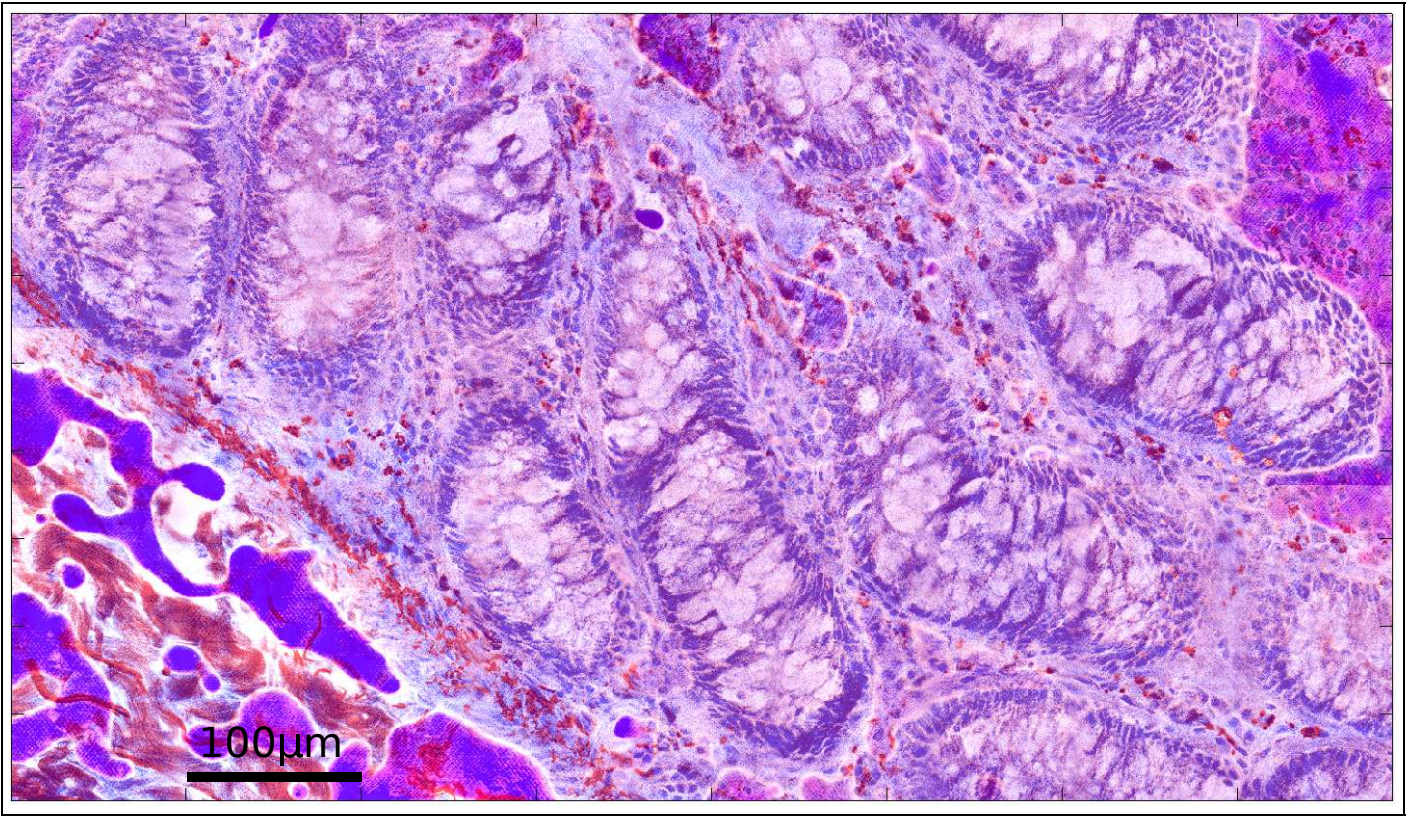}}
\caption{Frozen section of healthy human colon imaged using fast SRS and SHG. The spatial map of proteins (blue) and lipids (pink), were retrieved using the spectra of BSA and Oil for projection. The SHG signal, characteristic of collagen fibers is shown in red. Image total acquisition time 15 minutes.}
\label{fig:Histology}
\end{figure}

 Fast label free imaging of biopsy sections is a major application of coherent Raman imaging that can potentially revolutionize the field of histology by increasing diagnosis speed, lowering hospital costs, and improving patient care~\cite{Cicerone2018}.
Frozen sections of human cancer colon tissues were imaged with our fast SRS imaging platform, using the WB AOPDF with a pixel time of \SI{25}{\micro \second}. 8 by 10 adjacent fields of views (100um x 100um) were acquired separately and stitched together to reconstruct a wider image (fig.~\ref{fig:Histology}). The total acquisition time was 15 minutes.
Second harmonic generation at \SI{400}{\nano \meter} was recorded in the epi direction, to provide an additional contrast mechanism specific to collagen fibers. As previously for each pixel, the recorded hyperspectral images were projected on the spectra corresponding to pure BSA (protein) and oil (lipid) to highlight the nuclei and cell bodies, respectively. For rendering look up tables were adjusted to mimic eosin, saffron and haematoxylin staining~\cite{Orringer2017,Sarri2019}.

\section{Discussion}
Our SRS imaging platform acquires a full vibrational spectra over the [\SI{2850}{\per \centi \meter}, \SI{3050}{\per \centi \meter}] spectral range in \SI{12.5}{\micro \second}, this corresponds to the time during which the chirped pump and Stokes pulses overlapped, while the pump pulse delay is swept by the AOPDF. However the minimum total time per pixel is \SI{25}{\micro \second} when using the WB AOPDF corresponding to a duty cycle of 50 percent. This figure could be increased by designing a shorter AODPF delay line optimized for lower delay range and higher repetition rate. By applying dispersion with a grating pair, and using the AOPDF solely as a delay line, a \SI{80}{\kHz} pixel acquisition rate should be achievable.
In diluted samples such as biological tissues, image-wise averaging had to be performed to achieve signal to noise ratios (SNR) compatible with imaging (SNR > 10 dB). With our shot noise limited system, the SNR achieved on biological samples without averaging (pixel dwell time \SI{25}{\micro \second}) is on the order of 1, providing information on the associated SRS modulation.
Given a shot-noise limited system, with a lockin bandwidth $\Delta f = \SI{1.3}{\MHz}$, a probe beam average laser power $I_{p}$ of \SI{15}{\mW} at the wavelength $\lambda=\SI{800}{\nm}$, the SRS relative modulation $\beta=\frac{\Delta I}{I}$ associated with an SNR of 1 is given by~\cite{AudierNoise2019}:
\begin{equation}
    \beta = 4 \sqrt{\frac{h c \Delta f}{I_{p} \lambda \eta}} \approx 2\times 10^{-5}
\end{equation}
where $\eta$=0.8 is the detector quantum efficiency and $h$ the Planck constant.

The sensitivity of the system could be increased by red-shifting the SRS pump and Stokes wavelength.
Using near-infrared light such as \SI{940}{\nm} pump and \SI{1310}{\nm} Stokes, one may significantly increase the average laser power on the sample, while keeping photo-damage low.
Such near-IR scheme would allow for an increase of both laser powers by a factor of 2 to 3, which would be sufficient to improve the SNR by a factor of 10, therefore allowing for single SRS spectral measurements per pixel. 
We limited our investigation to the lipid band but the proposed approach can be easily extended to the fingerprint spectral range using appropriate AOPDF and large bandwidth tunable fs sources.  
Other improvements may include the use of spatial multiplexing, provided higher laser powers are available. Using multiple foci has been demonstrated as a method to scale up the acquisition rate~\cite{Heuke2018a} in SRS. Another possible strategy would be to use our fast spectral acquisition scheme while under-sampling the image through matrix completion scheme~\cite{Lin2018}.

\section{Conclusion}
We demonstrated fast hyperspectral SRS imaging by combining spectral focusing with an acousto-optic delay line working at \SI{40}{\kilo \hertz}. Our system is at the shot noise level for \SI{15}{\milli \watt} of laser probe power, and a bandwidth of \SI{1.3}{\MHz}.
The achieved resolution is \SI{22}{\per \cm} over a bandwidth of \SI{130}{\per \cm} full width at half maximum.
SRS spectra in the lipid band were acquired with \SI{12.5}{\micro \second} recordings per pixel corresponding to a duty cycle of 50 percent.
Samples containing up to 5 different chemical species have been imaged, and individual component maps were retrieved with no visible cross-talk.
Time-resolved imaging of a chemical transformation ($\delta$ Mannitol to $\beta$ Mannitol) was made possible, opening the door to other studies requiring fast chemical imaging.
Finally, the sensitivity of the system was sufficient to perform stimulated Raman imaging of human frozen tissue sections, demonstrating the applicability of this measurement system for label-free and live histology.

\section*{Funding Information}
We acknowledge financial support from the Centre National de la Recherche Scientifique (CNRS), Aix-Marseille University A*Midex (ANR-11-IDEX-0001-02) (A-M-AAP-ID-17-13-170228-15.22-RIGNEAULT), ANR grants France Bio Imaging (ANR-10-INSB-04-01) and France Life Imaging (ANR-11-INSB-0006) infrastructure networks and Plan cancer INSERM PC201508 and 18CP128-00.

\section*{Acknowledgments}
The authors would like to thank Barbara Sarri (Institut Fresnel) and Flora Poizat (Institut Paoli-Calmettes) for providing and helping with the frozen tissue sections.

\section*{Disclaimer}
Nicolas Forget has financial interest in FASTLITE.

\section*{Supplemental Documents}
Visualization 1: Mannitol dilution.

\bibliography{library} 
\bibliographystyle{ieeetr}

\end{document}